\documentstyle[11pt,newpasp,twoside,epsfig,oldlfont]{article}
\begin{document}


\newcommand{\be}{\begin{equation}}
\newcommand{\ee}{\end{equation}}
\newcommand{\bt}{\begin{table} \begin{center}}
\newcommand{\et}{\end{center} \end{table}}
\newcommand{\ba}{\begin{eqnarray}}
\newcommand{\ea}{\end{eqnarray}}
\newcommand{\ie}{{\it i.e.~}}
\newcommand{\eg}{{\it e.g.~}}

\newcommand{\citenp}{\cite}
\newcommand{\sch}{Schwarzchild~}
\newcommand{\BV}{Brunt-V\"ais\"al\"a~}
\newcommand{\mt}{\mathit}
\newcommand{\mr}{\mathrm}

\newcommand{\ms}{{\cal M}_\odot}
\newcommand{\ls}{{\cal L}_\odot}
\renewcommand{\sun}{\odot}
\def\ssb#1{\noindent{{\bf -- #1}}}
\newcommand{\tm}{{\tilde m}}
\newcommand{\Wt}{{\tilde W}}

\renewcommand{\thefootnote}{\fnsymbol{footnote}}


\def\lesssim{\mathrel{\mathchoice {\vcenter{\offinterlineskip\halign{\hfil
$\displaystyle##$\hfil\cr<\cr\sim\cr}}}
{\vcenter{\offinterlineskip\halign{\hfil$\textstyle##$\hfil\cr
<\cr\sim\cr}}}
{\vcenter{\offinterlineskip\halign{\hfil$\scriptstyle##$\hfil\cr
<\cr\sim\cr}}}
{\vcenter{\offinterlineskip\halign{\hfil$\scriptscriptstyle##$\hfil\cr
<\cr\sim\cr}}}}}
\def\gtrsim{\mathrel{\mathchoice {\vcenter{\offinterlineskip\halign{\hfil
$\displaystyle##$\hfil\cr>\cr\sim\cr}}}
{\vcenter{\offinterlineskip\halign{\hfil$\textstyle##$\hfil\cr
>\cr\sim\cr}}}
{\vcenter{\offinterlineskip\halign{\hfil$\scriptstyle##$\hfil\cr
>\cr\sim\cr}}}
{\vcenter{\offinterlineskip\halign{\hfil$\scriptscriptstyle##$\hfil\cr
>\cr\sim\cr}}}}}

\def\md{\dot{m}}
\def\Md{\dot{M}}
\def\L{{\cal L}}

\newcommand{\NLMC}{LMC~88~\#1}
\newcommand{\NFH}{FH~Ser}

\def\func{{\cal W}}


\bibliographystyle{apj}

 \title{Super Eddington Atmospheres and their Winds}


\author{Nir J. Shaviv}
\affil{Canadian Institute for Theoretical
        Astrophysics, University of Toronto \\ 60 St. George St.,
        Toronto, ON M5S 3H8, Canada}

\begin{abstract}

 We present a model for the steady state winds of super-Eddington
 systems. These radiatively driven winds are expected to be optically
 thick and clumpy as they arise from a porous atmosphere.  The model
 is then used to derive the mass loss of bright classical novae. The
 long duration super-Eddington outflows that are clearly observed in
 at least two cases (Nova LMC 1988 \#1 \& Nova FH Ser) are naturally
 explained. Moreover, the predicted mass loss and temperature
 evolution agree nicely with the observations, as do additional
 features. $\eta$ Car is then used to double check the theory which
 predicts the observed mass shed in the great eruption.

\end{abstract}

\section{Introduction}

 $\eta$ Car was super-Eddington (hereafter SE) during its 20 year long
 eruption (e.g., Davidson \& Humphreys 1997). Yet, its observed mass
 loss and velocity are inconsistent with a homogeneous solution for
 the wind (Shaviv 2000b). Basically, the sonic point obtained from the
 observed conditions necessarily has to reside too high in the
 atmosphere, at an optical depth of only $\sim 1$ to $\sim 300$, while
 the critical point in a homogeneous atmosphere necessarily has to
 reside at significantly deeper optical depths. The inconsistency
 arises because the sonic and critical points have to coincide in a
 steady state solution.

 A solution was proposed in which the atmosphere of $\eta$ Carinae is
 inhomogeneous, or porous (Shaviv 2000b). The inhomogeneity is a
 natural result of the instabilities of atmospheres that are close to
 the Eddington luminosity (Shaviv 2000a). The inhomogeneities, or
 ``porosity'', reduce the effective opacity and increase the effective
 Eddington luminosity (Shaviv 1998).  Here, we are interested in
 understanding the wind generated in cases in which the luminosity is
 SE.  To do so, it is advantageous to find a class of objects for
 which better data than for $\eta$ Car exists. One such class of
 objects is novae. Their advantage over other types is that their mass
 and luminosity is known better than all other SE objects, allowing a
 detailed analysis.

\section{The Fundamental Structure of Super-Eddington Winds (SEWs)}
\label{sec:wind}

 The above considerations lead us to propose the structure described
 in fig.~1 for a SE atmosphere and its SEW.  The most important point
 raised here is the identification of the location of the sonic point:
 If the flux corresponds to an Eddington parameter $\Gamma$, then the
 optical depth at which perturbations cannot reduce the effective
 Eddington parameter to unity, should scale as $\Gamma-1$. The reason
 is that the needed luminosity decrease should be proportional to the
 deviation of the actual luminosity from the Eddington one.  Namely,
 when close to the Eddington limit, a blob with the same geometry
 needs a smaller density fluctuation and with it a smaller change in
 the opacity, to reduce the effective opacity by the amount needed to
 become Eddington.  Thus, when closer to the Eddington limit, the
 sonic point can sit higher in the atmosphere.

\begin{figure}
\begin{minipage}{8in}
\hskip -2em
\begin{minipage}[b]{3.8in} 
 \caption{ The structure of a super Eddington atmosphere and the wind
 that it generates.  {\bf Region A}: A Convective envelope -- where
 the density is sufficiently high such that the excess flux above the
 Eddington luminosity is advected using convection. The radiative
 luminosity left is below the classical Eddington limit: $L_{\mr{rad}}
 < \L_{\mr{Edd}} < L_{\mr{tot}}$. Convection is always excited before
 the Eddington limit is reached. Thus, if the density is high enough
 and the total flux is SE, this region has to exist.  {\bf Region B}:
 A zone with lower densities, in which convection becomes
 inefficient. Instabilities render the atmosphere inhomogeneous, thus
 facilitating the transfer of flux without exerting as much force. The
 effective Eddington luminosity is larger than the classical Eddington
 luminosity: $\L_{\mr{Edd}} < L_{\mr{rad}} = L_{\mr{tot}} <
 \L_{\mr{eff}}$. $\eta$ Car has shown us that the existence of this
 region allows for the }
\end{minipage}
\hskip 0.1in
\epsfig{file=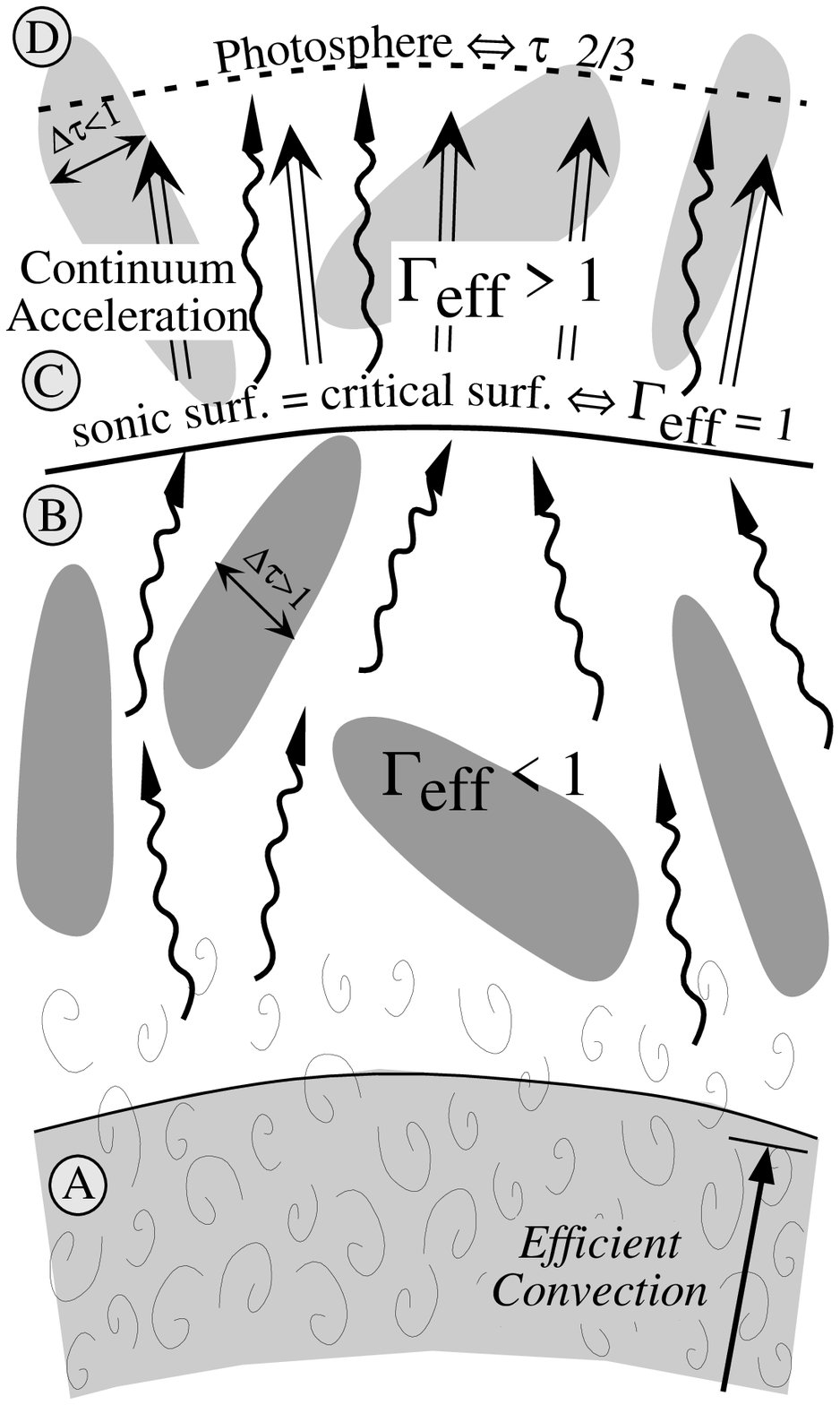,width=1.8in}
\end{minipage}
\begin{minipage}{5.25in}
      steady state outflow during its 20 year long eruption (Shaviv
      2000b).  {\bf Region C}: A region in which the effect of the
      inhomogeneities disappears and the luminosity is again SE. When
      perturbations arising from the instabilities, which are expected
      to be of order the scale height in size, become transparent, the
      effective opacity tends to the microscopic value and the
      effective Eddington limit tends to the classical value. At the
      transition between (B) and (C), the effective Eddington is equal
      to the total luminosity. This critical point is also the sonic
      point in a steady state wind. Above the transition surface,
      $L_{\mr{tot}} > \L_{\mr{eff}} \rightarrow \L_{\mr{Edd}}$ and we
      have an optically thick super-sonic wind.  {\bf Region D}: The
      photosphere and above. Since the wind is thick, region (C) is
      wide.
\end{minipage}
\label{fig:structure}
\vskip 0cm
\end{figure}

By writing the expression for the scale height, which is expected to
be of order the typical size of the perturbations, one obtains the
mass loss:
\vskip -0.4cm
\be
 \Md = 4\pi R_\star^2 f \rho_{\star} v_s = \func(\Gamma)
 {L_{\mr{tot}}-\L_{\mr{Edd}} \over c v_s},
\label{eq:masslossrate}
\ee
\vskip -0.2cm
\noindent
 where $1-f$ is the covering factor of the dense blobs and
 $\func(\Gamma)$ is a dimensionless function that is expected to be
 almost a constant of order of unity, or perhaps somewhat larger,
 depending on the efficiency at which the effective opacity can be
 reduced (the smaller the efficiency, the larger will $\func$ be). It
 is only with more elaborate simulations or with more accurate
 observations, that a more accurate functional form can be
 deduced. For the meantime, we settle with this simplifying yet
 reasonable assumption.  We will show in the rest of the paper that
 eq.~1 provides an explanation to the mass loss from bright classical
 novae as well as from $\eta$ Car and allows us to connect between the
 observed luminosity and mass loss rate, a relation which hitherto did
 not exist for SE systems.

 Before we compare to observations, we need to consider that the winds
 generated are heavy and a significant part of the radiation can be
 used to drive the wind up the potential well and to accelerated
 it. We use the relations of Owocki \& Gayley (1997) to relate the
 parameters important for the wind at its base (e.g., $\Gamma_\star$,
 $v_{esc}$) to the parameters observed at infinity (e.g.,
 $\Gamma_\infty$, $v_{\mr{esc}}$). An example can be found in fig.~2 which
 depicts the observed parameters as a function of $\Gamma_\star$ and
 $\tilde{\func} \equiv v_{\mr{esc}} / (2 v_s c) \func$.

\section{Application of the SEW theory}
\label{sec:novae}

We now proceed to apply the SEW theory to systems that clearly
exhibited SE outflows over dynamically long durations. We then
continue with general predictions pertaining to novae and SEWs.

\ssb{The wind parameter:}
 Each of the three SE objects was studied in a different manner since
 the observational data was different. \NFH~has measurements of the
 time evolution of the bolometric flux, color temperature and velocity
 at infinity (Friedjung 1989). The data for \NLMC~(Schwarz 1998) do
 not have proper measurements of the velocity evolution, but the
 temperatures given are the effective ones. $\eta$ Car is a markably
 different system than that of novae and since it doesn't have proper
 measurements of the evolution of temperature and luminosity during
 its great eruption, we just take the estimates given by Davidson
 (1999) for the integrated mass loss and energy radiated during the
 eruption.  Using the SEW theory we find the following values for the
 wind parameter: $ \func = 2.8 \pm 1.4 ,(7.5\pm 4)/\delta^{3/4} , 4.5
 \pm 3.3$ for Novae LMC 88\#1, FH Ser, and $\eta$ Car
 respectively. $\delta^2\equiv{\left< \rho^2 \right> / \left< \rho
 \right>^2}$ is the clumpiness parameter. It enters the analysis of
 \NFH~when translating the color temperature into a mass loss rate. As
 expected, we find that all $\func$'s obtained are consistent with
 each other and the theory provided that there is some clumping in the
 wind. Clumping is however predicted since SEWs originate from
 inhomogeneous atmospheres. The amount of clumping needed is similar
 to that already observed in WR winds (e.g., St.-Louis et al.~1993).

\ssb{Temperature evolution:}
 In the case of FH Ser, where good data for the evolution of the
 velocity are present, a better prediction for the evolution of the
 temperature can be made using the observed luminosity . The agreement
 between the predicted temperature and the observed temperature is
 seen in fig.~3.

\begin{figure}
\begin{minipage}{6in}
\hskip -2.0em
\begin{minipage}[t]{3.25in}
\hskip 1.8em
\epsfig{file=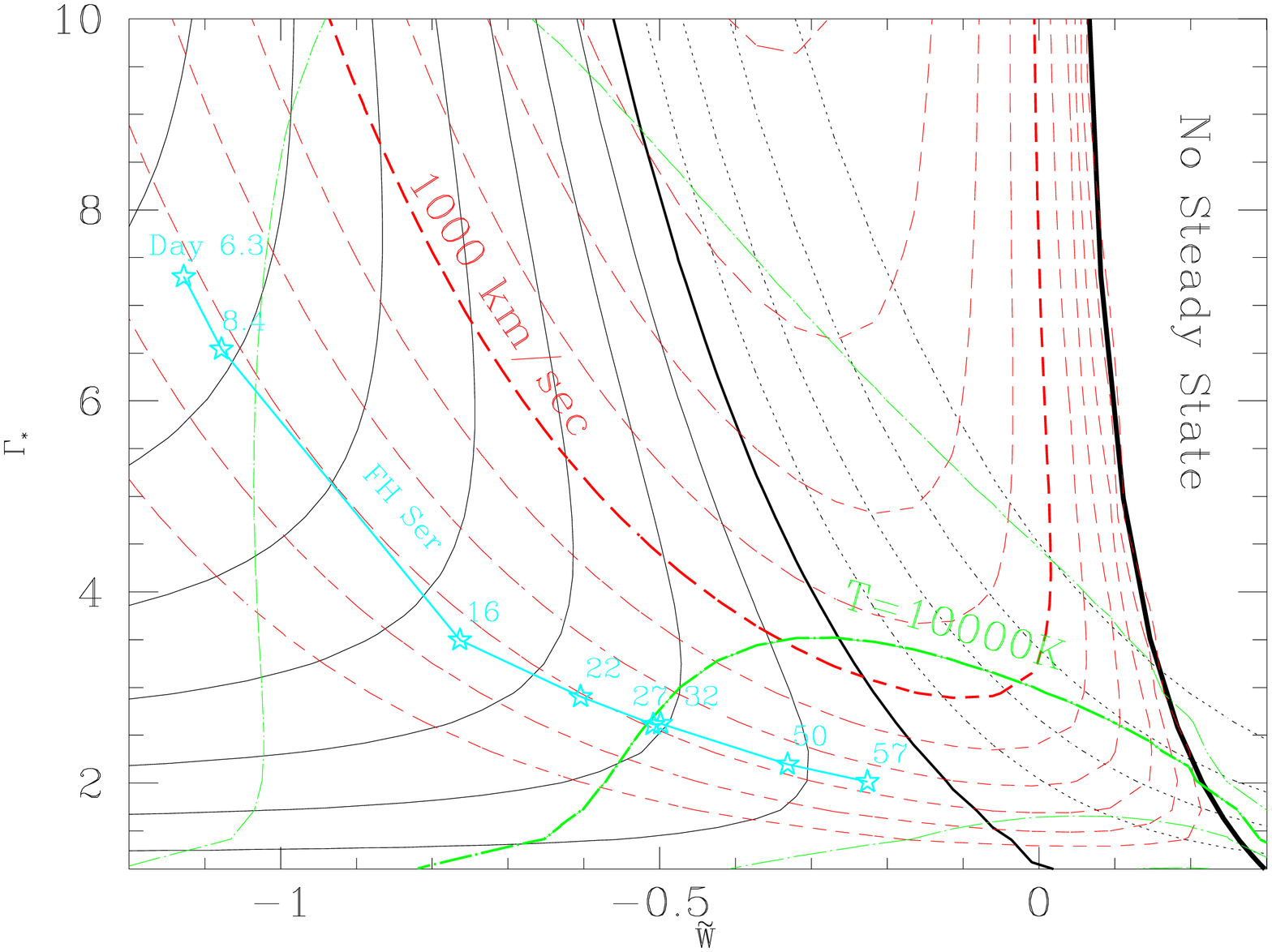,angle=-90,width=2.5in}

\caption{
  The evolution of FH Ser in the $\Gamma_\star$ - $\Wt$ plane.  The
  lines are iso-contours of $\Gamma_{\infty}$ (the solid lines for
  $\Gamma_{\infty} \geq 1 $ are spaced at 0.2 dex, and dotted spaced
  linearly at 0.2 intervals for $\Gamma_{\infty}<1$), $v_{\infty}$
  (short-long dashed lines spaced at intervals of 100 km/sec, higher
  are larger) and the color temperature of the photosphere
  $T_{\mr{ph}}$ (dot-dashed lines spaced at $2500^\circ{\mr{K}}$,
  lower is hotter).}
\end{minipage}
\hskip -2em
\begin{minipage}[t]{2.85in}
\hskip 3mm
\epsfig{file=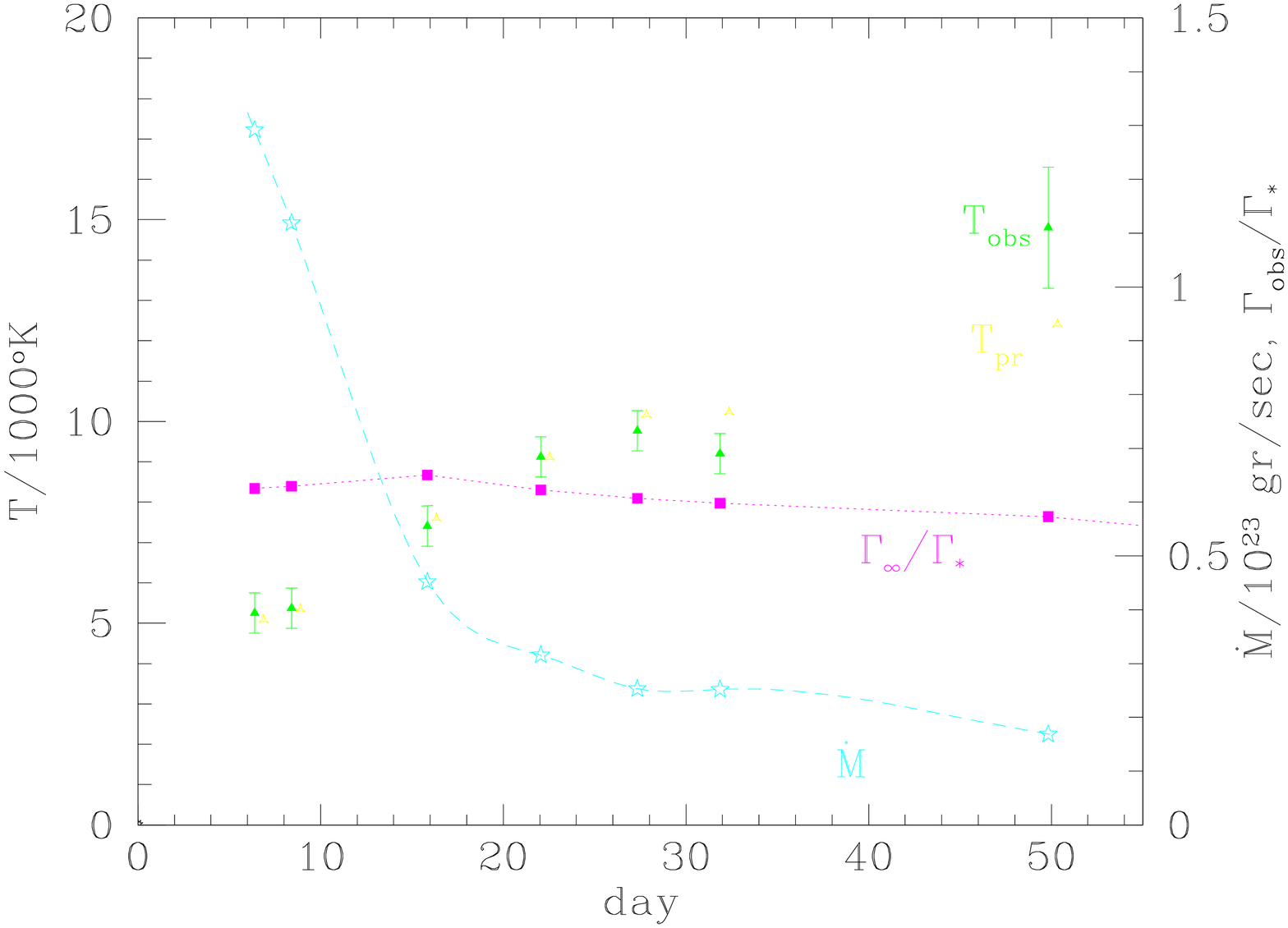,angle=-90,width=2.6in}

\caption{
  The observed temperature behavior $T_{\mr{obs}}$ of \NFH~as compared
  with the predicted temperature behavior $T_{\mr{pr}}$ from the
  observed luminosity and wind model, for the best fit model assuming
  $M_{\mr{WD}}=0.8 M_\sun$ and $X=0.7$. Additional plots are of the
  mass loss and the ratio of $\Gamma_{\infty}$ to $\Gamma_\star$.  }
\label{fig:FHTemp}
\end{minipage}
\end{minipage}
\end{figure}

\ssb{Super Eddington fluxes:}
  In a homogeneous atmosphere, one expects the steady state structure
  to relax into a sub-Eddington state, as is given by the core-mass
  luminosity relation (Paczynski 1970). The fact that a reduced
  effective opacity is obtained for an inhomogeneous system, implies
  that the ``saturation'' luminosity should be the modified Eddington
  luminosity, which of course is larger than the classical Eddington
  one.

\ssb{The ``Transition Phase'':}
 One of the seemingly odd behavior that is displayed by a large
 majority of all the classical nova eruptions is a transition
 phase. If it appears, it starts once the visual magnitude has decayed
 by 3 to 4 magnitudes. During the transition phase, the light curve
 can display strong deepening, quasi-periodic oscillations, erratic
 changes or other complicated behavior.  The SEW theory can naturally
 explain the transition phase. If we look at the $\Gamma_0$ - $\Wt$
 trajectory of \NFH~in fig.~2 (which had a transition phase), one
 cannot avoid the extrapolation of the trajectory into the zone of
 ``no steady state configuration''. In this zone, the wind is too
 thick for the radiation field to push to infinity. The wind then
 naturally stagnates and it does not allow a steady state. Non trivial
 2D or 3D flows that must result could potentially yield the non
 trivial variability observed.  The phase is expected to end once the
 luminosity at the base of the wind falls below the Eddington limit,
 shutting off the SEW, at which point the ``naked'' white dwarf should
 emerge.

\ssb{General Mass Loss of Novae:}
 We can use the trends observed for the nova population in general to
 predict an average mass-loss. To do so, we create a template nova as
 a function of white dwarf mass and explore its properties. Its
 predicted mass loss is then compared with the observed integrated
 mass loss. Clearly, we expect the theoretical prediction to provide
 the guide line to which the {\em average} behavior should be compared
 to.  Using general relations between average observed novae
 parameters, we can write down the peak bolometric luminosity of the
 nova as a function of WD mass. We assume that the transition phase
 corresponds to the nova's base bolometric luminosity crossing the
 Eddington limit. If we further assume that it decays exponentially
 (or linearly), we can calculate the total mass loss by integrating
 eq.~1. The results are plotted in fig.~4 together with the observed
 determinations of ejecta masses. Evidently, the SE bolometric
 behavior of classical novae agrees well with the observed mass loss
 if we use the SEW theory described here. Note that the standard
 simulations of TNRs using a {\em homogeneous} atmosphere predict a
 mass loss that is an order of magnitude smaller for the large WD
 masses.

\begin{figure}[tbh]
\begin{minipage}{6in}
\hskip -2em
\begin{minipage}[t]{2.45in}
\caption{ Nova ejected mass vs.~WD mass.  The solid line is obtained
 by the wind model assuming $\func=2.8$, the long dashed line when $X$
 is changed from 0.7 to 0.3, the upper (and lower) short dashed lines
 when taking a value for $\func$ which is higher (lower) by 1.5, the
 dash-dotted line when a linear decay is assumed, and the dotted when
 taking into account the natural scatter in the $t_3$-$M_{\mr{WD}}$
 relation. }
\end{minipage}
\hskip 1mm
\begin{minipage}[t]{2.35in}
\vskip 5mm
\epsfig{file=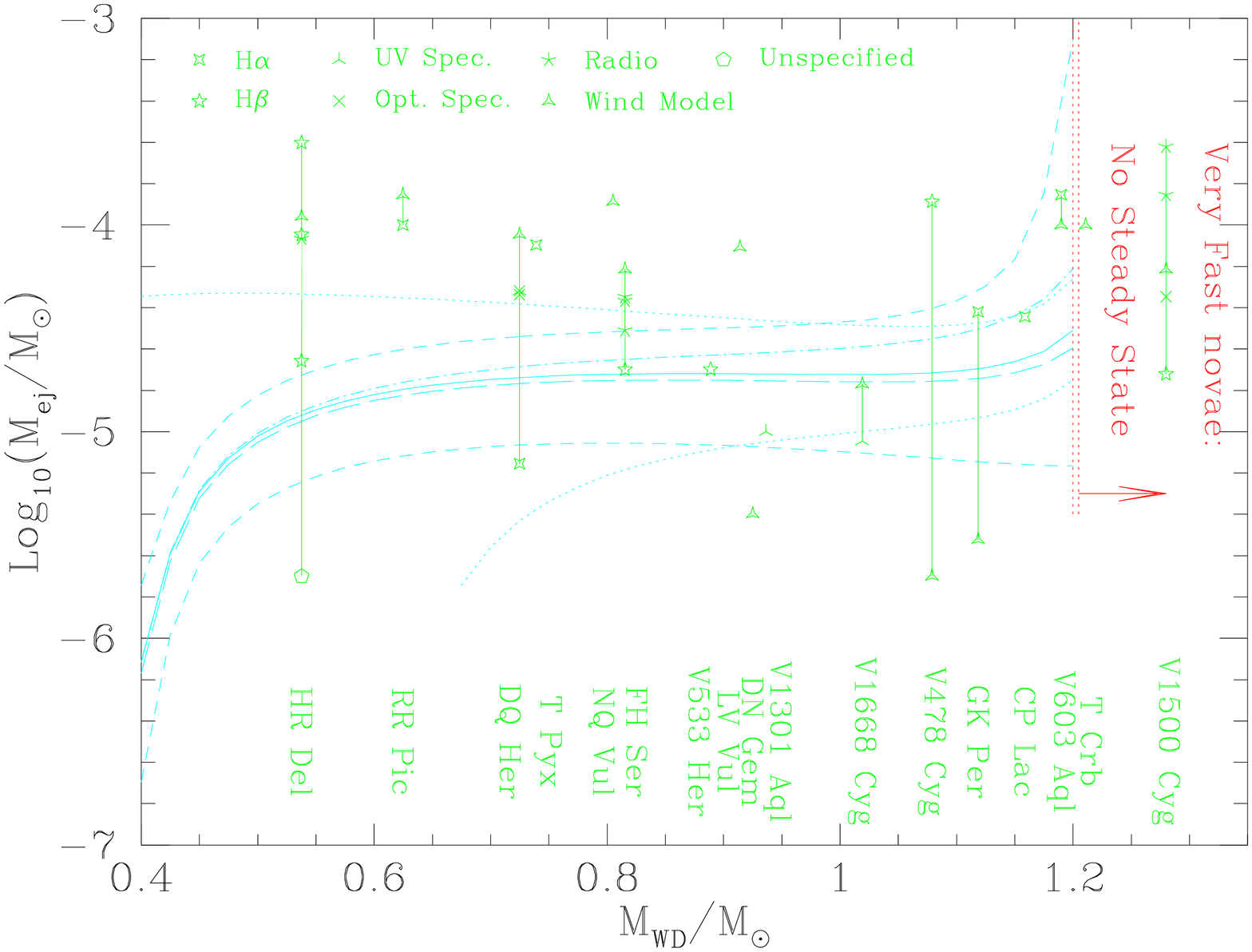,width=2.3in,angle=-90}
\label{fig:masslossmass}
\end{minipage}
\end{minipage}
\end{figure}

\ssb{A ``Constant Bolometric Flux'':}
 Some novae appear to have a plateau in their bolometric flux.  If we
 look at fig.~2 and note that lines of constant radii are almost
 vertical, we see that for moderately ``loaded'' winds and
 $\Gamma_\star$ of a few, an evolution in which the temperature
 increases but the apparent luminosity remains constant is possible if
 the radius does not decrease dramatically.  Under such conditions,
 the base luminosity of the wind does fall off to get a higher
 temperature, however, the lower mass loss predicted implies that less
 energy is needed to accelerate the material to infinity and so a
 larger fraction of the base luminosity remains after the wind has
 been accelerated, yielding an almost constant $\Gamma_\infty$.

\ssb{Clumpiness of the wind:}
 Since the atmospheres generating the SEWs are necessarily clumpy, the
 winds will be clumpy as well. By estimating the typical perturbation
 size in the unstable atmosphere, one can estimate the size of the
 blobs. In the case of novae and the instabilities found by Shaviv
 (2000a), the expected peak spherical harmonic of the clumpy wind is
 $\ell \lesssim 1000$.

\section{Discussion \& Summary}

 We have tried to present the following coherent picture which
 explains the observed existence of SE atmospheres: Homogeneous
 atmospheres becomes inhomogeneous as the the radiative flux
 approaches the Eddington limit. This is due to a plethora of
 instabilities. The particular governing instability depends on the
 details of the atmosphere. Due to the inhomogeneity, the effective
 opacity is then reduced as it is easier for the radiation to escape;
 consequently, the effective Eddington limit increases.  SE
 configurations are now possible because the bulk of the atmosphere is
 effectively sub-Eddington. Very deep layers advect the excess total
 luminosity above Eddington by convection. Higher in the atmosphere,
 where convection is inefficient, the Eddington limit is effectively
 increased due to the reduced effective opacity. The top part of the
 atmosphere, where perturbations of order of the scale height become
 optically thin, has however to remain SE. Thus, these layers are
 pushed off by a continuum driven wind.

 By identifying the location of the critical point of the outflow, one
 can obtain a mass-loss luminosity relation. The relation, given by
 eq.~1 is the main result of the paper. The relation has a universal
 dimensionless parameter which should be of order or somewhat larger
 than unity.

 To check the result, we analyzed 2 novae and the star $\eta$
 Car. Although the two types of systems are markably different, as
 they have masses, luminosities and massloss rates which differ by
 orders of magnitude, the wind mass loss and the wind parameter are
 found to be in good agreement with the theoretical
 expectation. Moreover, the results are completely consistent with
 clumping --- a natural prediction of SEWs since the atmospheric
 layers beneath the sonic point are predicted to be inhomogeneous.

 We also identify the occurrence of the ``transition phase'' observed
 in a majority of the novae with the advance of the atmosphere into
 the ``no steady state region''.  As the nova explosion progresses,
 its luminosity and radius decline. However, if the radius decreases
 too quickly, at some point the SEW predicted will be too heavy for
 the luminosity at the base to push to infinity. No steady state will
 then exist. The inconsistency might explain the strange behaviors
 observed in the transition phase of different novae.

 The wind model presented is by no means a complete theory for novae
 since it cannot predict {\em ab initio} the luminosity at the base of
 the wind. To obtain the latter, one needs to solve for the complete
 evolution of a nova taking into account the porosity and lowered
 opacity in the nova atmosphere. One expects that the lowered opacity
 increases the luminosity obtained in the core-mass luminosity
 relation, and super-Eddington values therefore arise naturally.

\end{document}